\documentstyle[prl,aps]{revtex}
\begin{document}

\draft

\title{An Alternative to See-Saw}

\author{J. I. Silva-Marcos\cite{juca}}
\address{NIKHEF, Kruislaan 409, 1098 SJ Amsterdam, The Netherlands}

\maketitle

\begin{abstract}

We give a new mechanism for generating very small, and almost degenerate,
neutrino masses, without resorting to the see-saw mechanism or unnatural
small Yukawa couplings. It requires the existence of at least 4 families
with an almost democratic structure for the Yukawa couplings. 
It is also proven that this structure can account for 
large lepton mixings of the three
light leptons.
\end{abstract}

As is well known, the famous see-saw mechanism $\cite{ref1}$ is particularly
attractive for explaining the smallness of the three neutrino masses.
However, another mechanism might be responsible for very small neutrino
masses. Within the framework of 4 families $\cite{ref1a}$, an almost
democratic structure for the Yukawa couplings is sufficient to explain the
smallness of the masses of the three known neutrinos. One can see this by
considering the democratic limit for the neutrino mass matrix, 
\begin{equation}
\label{eq1}m_\nu =\frac{\mu _\nu }4\left[ 
\begin{array}{cccc}
1 & 1 & 1 & 1 \\ 
1 & 1 & 1 & 1 \\ 
1 & 1 & 1 & 1 \\ 
1 & 1 & 1 & 1 
\end{array}
\right] \quad 
\end{equation}
Obviously, the eigenvalue spectrum of this mass matrix consists of three
neutrinos with zero mass and one massive heavy neutrino, 
$(0,0,0,\mu _\nu )$. The democratic structure for the neutrino 
mass matrix is very appealing.
On the one hand, all Yukawa couplings can be as large as any of the other
fermion Yukawa couplings, while, on the other hand, all, but one,
eigenvalues of the mass matrix are zero, or practically zero in the case of
an almost democratic structure.  Furthermore, democratic mass matrices 
are related to exact $S_n$ family symmetries.
The existence of three very light neutrinos,
leads us to consider, at least, 4 families and an almost democratic
structure. Thus, we can explain very small neutrino masses without resorting
to the see-saw mechanism or unnatural small Yukawa couplings.

The universal strength for Yukawa couplings (USY) $\cite{ref2}$ is a natural
framework to study small deviations from the democratic limit $\cite{ref3}$.
It was proven $\cite{ref4}$, that for three generations, USY can account for
almost degenerate neutrinos $\cite{ref5}$ and for the strong hierarchy in
the charged lepton sector. In this paper we shall further generalize these
results to 4 generations of Dirac neutrinos. With regard to lepton mixing,
we shall consider here only cases where the mixing (mainly) occurs for the
light leptons. Thus the mixings of the fourth heavy neutrino with the three
light charged leptons and the mixings of the fourth heavy charged lepton
with the three light neutrinos will be very small.

As a first step, we analyze two preliminary limit situations, one for the
neutrino and the other for the charged leptons, which represent small
deviations from the exact democratic limit in Eq.(\ref{eq1}), and which
serve as simplifications of a more realistic scenario given further on.

For the neutrinos, in this first step, the three light neutrinos acquire a
mass but maintain their degeneracy. We achieve this by allowing a very small
phase $\alpha $ for the diagonal elements of the neutrino mass matrix, such
that the democratic limit of Eq.(\ref{eq1}) becomes, 
\begin{equation}
\label{eq2}m_\nu =\frac{\mu _\nu }4\left[ 
\begin{array}{cccc}
e^{i\alpha } & 1 & 1 & 1 \\ 
1 & e^{i\alpha } & 1 & 1 \\ 
1 & 1 & e^{i\alpha } & 1 \\ 
1 & 1 & 1 & e^{i\alpha } 
\end{array}
\right] \quad 
\end{equation}
As can be readily verified, the mass spectrum of this neutrino mass matrix
consists of three light neutrinos with equal mass, and one heavy neutrino: 
\begin{equation}
\label{eq3}
\begin{array}{lll}
m_{\nu _i}=\frac 12\ \mu _\nu \ \left| \sin (\frac \alpha 2)\right| \quad
,\quad i=1,2,3 & \ ;\qquad & m_{\nu _4}=\mu _\nu \ \sqrt{1-\frac 34\sin
^2(\frac \alpha 2)} 
\end{array}
\end{equation}
It is clear that one can choose the mass of the light neutrinos as small as
one wants by taking the phase $\alpha $ to be very small. Thus the 
breaking of the $S_4$ symmetries will be very small. The
diagonalization of this neutrino mass matrix is done with:
\begin{equation}
\label{eq4}F_\nu =\left[ 
\begin{array}{cccc}
\frac 1{\sqrt{2}} & \frac{-1}{\sqrt{6}} & \frac{-1}{\sqrt{12}} & \frac 12 \\ 
0 & \frac 2{\sqrt{6}} & \frac{-1}{\sqrt{12}} & \frac 12 \\ 
0 & 0 & \frac 3{\sqrt{12}} & \frac 12 \\ 
\frac{-1}{\sqrt{2}} & \frac{-1}{\sqrt{6}} & \frac{-1}{\sqrt{12}} & \frac 12 
\end{array}
\right] \quad 
\end{equation}
One must realize that, at this stage, that there is much freedom in the
choice of the first three column vectors of the unitary matrix of Eq.(\ref
{eq4}), because of the degeneracy of the three light eigenvalues. Therefore,
these three vectors are (for now) given, up to modulo $3\times 3$ unitary
rotations, which, of course, will be fixed when the light neutrino
degeneracy is lifted.

With regard to the charged leptons, we consider in this first step the limit
when the electron and muon are massless while the tau acquires a mass (as
well as the heavy fourth heavy charged lepton). One can accomplish this with
just one phase $r$, 
\begin{equation}
\label{eq5}m_{{\ell }}=\frac{\mu _{{\ell }}}4\left[ 
\begin{array}{cccc}
1 & 1 & 1 & 1 \\ 
1 & 1 & 1 & 1 \\ 
1 & 1 & 1 & 1 \\ 
1 & 1 & 1 & e^{ir} 
\end{array}
\right] \quad 
\end{equation}
and one gets the following masses for the for the tau and the heavy charged
lepton, 
\begin{equation}
\label{eq6}
\begin{array}{lll}
m_\tau =\frac 3{4\sqrt{2}}\ \ \mu _{{\ell }}\ \frac{\left| \sin (\frac
r2)\right| }{\sqrt{1+\left( 1-\frac 9{16}\sin ^2(\frac r2)\right) ^{1/2}}}%
\quad & ; & m_{{\ell }_4}=\frac 1{\sqrt{2}}\ \ \mu _{{\ell }}\ \sqrt{%
1+\left( 1-\frac 9{16}\sin ^2(\frac r2)\right) ^{1/2}} 
\end{array}
\end{equation}
From Eq.(\ref{eq6}) one finds, in first order in $r$, that $m_\tau /m_{{\ell 
}_4}=3|r|/16$, which for a value of $m_{{\ell }_4}=5\ TeV$ would fix $%
r=1.9\times 10^{-3}$. The diagonalization of $m_{{\ell }}$ in Eq.(\ref{eq5}%
) requires a matrix

\begin{equation}
\label{eq7}F_{{\ell }}=\left[ 
\begin{array}{cccc}
\frac 1{\sqrt{2}} & \frac{-1}{\sqrt{6}} & \frac{-1}{\sqrt{12}} & \frac 12 \\ 
\frac{-1}{\sqrt{2}} & \frac{-1}{\sqrt{6}} & \frac{-1}{\sqrt{12}} & \frac 12
\\ 
0 & \frac 2{\sqrt{6}} & \frac{-1}{\sqrt{12}} & \frac 12 \\ 
0 & 0 & \frac 3{\sqrt{12}} & \frac 12 
\end{array}
\right] \quad 
\end{equation}
but again, here one must bear in mind that the first two charged lepton
masses are still zero, and that therefore, the first two column vectors of $%
F_{{\ell }}$ are, in principal, given up to modulo unitary $2\times 2$
rotations, which shall be fixed when the electron and the muon acquire a
mass. The unitary matrix $F_{{\ell }}$ of Eq.(\ref{eq7}) is not the full
matrix which diagonalizes the charged lepton mass matrix of Eq.(\ref{eq5})
because of mixing between the tau and the heavy charged lepton. However, in
first order, it is a very good approximation. It can be readily verified
that the mixing between the tau and the charged lepton is of the order of $%
o(m_\tau /m_{{\ell }_4})$. The lepton mixing matrix is given (in first
order) by,

\begin{equation}
\label{eq8}V_{\text{mixing}}^{\text{lepton}}=F_{{\ell }}^{\dagger }\cdot F_{{%
\nu }}=\left[ 
\begin{array}{cccc}
\frac 12 & \frac{-\sqrt{3}}2 & 0 & 0 \\ 
\frac{-\sqrt{3}}6 & \frac{-1}6 & \frac{4\sqrt{2}}6 & 0 \\ 
\frac{-\sqrt{6}}3 & \frac{-\sqrt{2}}3 & \frac{-1}3 & 0 \\ 
0 & 0 & 0 & 1 
\end{array}
\right] 
\end{equation}
One sees that relevant mixing between the leptons only occurs for the light
species.

The limit cases which we have considered so far for the neutrino as well as
the charged lepton mass matrices and diagonalizations are crucial. In
essence, they contain the first order approximations of a realistic example
which we present now, and where the neutrino degeneracy is lifted and the
electron and muon acquire mass. As an example of a realistic ansatz we
propose: 
\begin{equation}
\label{eq9}
\begin{array}{lll}
m_\nu =\frac{\mu _\nu }4\left[ 
\begin{array}{cccc}
e^{i\alpha } & 1 & 1 & 1 \\ 
1 & e^{i\alpha (1+\varepsilon )} & 1 & 1 \\ 
1 & 1 & e^{i\alpha (1+\delta )} & 1 \\ 
1 & 1 & 1 & e^{i\alpha } 
\end{array}
\right] & \quad ;\quad & m_{{\ell }}=\frac{\mu _{{\ell }}}4\left[ 
\begin{array}{cccc}
1 & 1 & 1 & 1 \\ 
1 & e^{ip} & 1 & 1 \\ 
1 & 1 & e^{iq} & 1 \\ 
1 & 1 & 1 & e^{ir} 
\end{array}
\right] 
\end{array}
\end{equation}
The slight deviation from the limit case of the neutrino mass matrix in Eq.(%
\ref{eq2}), induced by $\delta ,\varepsilon \neq 0$, will generate the
neutrino mass differences $\Delta m_{32}^2$ and $\Delta m_{21}^2$. One finds
that $\delta =o(\Delta m_{32}^2/m_{\nu _3}^2)$ and $\varepsilon =o(\Delta
m_{21}^2/m_{\nu _3}^2)$. This mass matrix has one exact light 
neutrino mass eigenvalue as given in Eq.(\ref{eq3}). 
For the charged lepton mass matrix, $q\neq 0$ and $%
p\neq 0$ will generate masses respectively for the muon and electron. One
has $q=o(m_\mu /m_{{\ell }_4})$ and $p=o(m_e/m_{{\ell }_4})$ $\cite{ref6}$.
The realistic ansatz of Eq.(\ref{eq9}) and its choice of phases, will fix
the explained arbitrary unitary rotations associated with the degeneracy of
the three light neutrinos and the zero masses of the electron and muon,
because of the explicit lifting of that degeneracy and the non-zero masses.
However, this choice of phases is such that these rotations are equal to
the identity. Therefore, in first order, we shall have precisely the same
diagonalization matrices as in Eqs.(\ref{eq4}, \ref{eq7}) and a lepton
mixing matrix as in Eq.(\ref{eq8}). To complete our ansatz, we give an exact
numerical example with input:

\begin{equation}
\label{eq10}
\begin{array}{lll}
\alpha =2\times 10^{-8} & ;\ \delta =2\times 10^{-3} & ;\ \varepsilon
=3\times 10^{-6} \\ 
r=1.9\times 10^{-3} & ;\ q=1.3\times 10^{-4} & ;\ p=8.2\times 10^{-7} 
\end{array}
\end{equation}
which corresponds with the following output: 
\begin{equation}
\label{eq10a}
\begin{array}{lll}
\Delta m_{21}^2=9\times 10^{-6}\ eV^2 & ;\ \Delta m_{32}^2=6.7\times
10^{-3}\ eV^2\  & ;\ m_{\nu _3}=1.5\ eV \\ 
m_e=511\ KeV\  & ;\ m_\mu =105.7\ MeV & ;\ m_\tau =1777\ MeV 
\end{array}
\end{equation}
In addition, there will be two heavy leptons associated with (the large
values of) $\mu _\nu $ and $\mu _{{\ell }}$. One obtains $\mu _\nu =\ m_{\nu
_4}=300\ GeV\ $and $\mu _{{\ell }}=m_{{\ell }_4}=5\ TeV$ . The mass of the
heavy neutrino is far above the LEP bound $M_Z/2$. The lepton mixing matrix
will also have contributions of higher orders, 
\begin{equation}
\label{eq11}\left| V_{\text{mixing}}^{\text{lepton}}\right| =\left[ 
\begin{array}{cccc}
0.501 & 0.865 & 1.1\times 10^{-3} & 1.4\times 10^{-7} \\ 
0.305 & 0.176 & 0.936 & 1.7\times 10^{-5} \\ 
0.810 & 0.469 & 0.353 & 2.0\times 10^{-4} \\ 
1.7\times 10^{-4} & 9.6\times 10^{-5} & 5.4\times 10^{-5} & 1. 
\end{array}
\right] 
\end{equation}

To summarize, we have given a new mechanism for generating very small, and
almost degenerate, neutrino masses, without resorting to the see-saw
mechanism or unnatural small Yukawa couplings. This new mechanism requires
the existence of at least 4 families$\cite{ref7}$. Within the framework of
the universal strength for Yukawa couplings one only has very small phases,
which may be associated with almost exact $S_4$ symmetries. It was also proven that
this structure for the Yukawa couplings can account for large lepton mixing
of the three light leptons.

This work was partially supported by the Funda\c c\~ao para a Ci\^encia e
Tecnologia of the Portuguese Ministry of Science and Technology.

\end{document}